\documentclass[runningheads]{llncs}

\pdfoutput=1

\usepackage{graphicx}
\usepackage{booktabs}
\usepackage{float}
\usepackage{fancyvrb}
\usepackage{amsmath,amssymb,amsfonts}
\usepackage{wasysym}
\usepackage{multirow}
\usepackage{makecell}
\usepackage{listings}
\usepackage{subcaption}
\usepackage{xcolor}
\usepackage{listings}

\newcolumntype{L}[1]{>{\raggedright\let\newline\\\arraybackslash\hspace{0pt}}m{#1}}
\newcolumntype{C}[1]{>{\centering\let\newline\\\arraybackslash\hspace{0pt}}m{#1}}
\newcolumntype{R}[1]{>{\raggedleft\let\newline\\\arraybackslash\hspace{0pt}}m{#1}}

\newcommand{\etal}{et al.~}
\newcommand{\ie}{i.e.,~}
\newcommand{\eg}{e.g.,~}

\newcommand{\figname}{Fig.~}

\AtBeginDocument{%
  \counterwithout{lstlisting}{chapter}%
}
\usepackage{listings}

\definecolor{editorOrange}{cmyk}{0, 0.8, 1, 0}

\lstdefinelanguage{DatalogFacts}{
  keywords=[1]{call, used_in_condition, transaction, block, erc20_transfer, opcode, uses, arithmetic, storage, condition, selfdestruct, data_flow, follows},
  keywordstyle=[1]\color{blue}\bfseries,
  keywords=[2]{type, decl, number, symbol},
  keywordstyle=[2]\color{violet}\bfseries,
  keywords=[3]{Address, Opcode, Value},
  keywordstyle=[3]\color{teal}\bfseries
}

\lstdefinelanguage{DatalogQuery}{
  keywords=[1]{UnhandledException, Reentrancy, IntegerOverflow, ParityWalletHack1, ParityWalletHack2, call, used_in_condition, transaction, block, ShortAddress, erc20_transfer, opcode, uses, arithmetic, storage, condition, throw, selfdestruct, error, data_flow, follows},
  keywordstyle=[1]\color{blue}\bfseries,
  keywords=[2]{hash, hash1, hash2, timestamp, step, step1, step2, step3, step4, depth1, depth2, index, index1, index2, block1, block2, contract, destination, operand1, operand2, arithmetic_res, evm_res, caller, callee, amount, from, to, input, input1, input2, type, decl, number, symbol, id, branch},
  keywordstyle=[2]\color{violet}\bfseries,
  keywords=[3]{substr, match, strlen},
  keywordstyle=[3]\color{teal}\bfseries,
}

\begin{document}
\title{The Eye of \textsc{Horus}: Spotting and Analyzing Attacks on Ethereum Smart Contracts}
\titlerunning{Spotting and Analyzing Attacks on Ethereum Smart Contracts}
\author{
Christof Ferreira Torres\inst{1} \and
Antonio Iannillo\inst{1} \and
Arthur Gervais\inst{2} \and
Radu State\inst{1}
}
\authorrunning{C. Ferreira Torres et al.}
\institute{
SnT, University of Luxembourg
\and
Imperial College London
}
\maketitle              %
\begin{abstract}
In recent years, Ethereum gained tremendously in popularity, growing from a daily transaction average of 10K in January 2016 to an average of 500K in January 2020.
Similarly, smart contracts began to carry more value, making them appealing targets for attackers.
As a result, they started to become victims of attacks, costing millions of dollars.
In response to these attacks, both academia and industry proposed a plethora of tools to scan smart contracts for vulnerabilities before deploying them on the blockchain.
However, most of these tools solely focus on detecting vulnerabilities and not attacks, let alone quantifying or tracing the number of stolen assets.
In this paper, we present \textsc{Horus}, a framework that empowers the automated detection and investigation of smart contract attacks based on logic-driven and graph-driven analysis of transactions.
\textsc{Horus} provides quick means to quantify and trace the flow of stolen assets across the Ethereum blockchain.
We perform a large-scale analysis of all the smart contracts deployed on Ethereum until May 2020.
We identified 1,888 attacked smart contracts and 8,095 adversarial transactions in the wild.
Our investigation shows that the number of attacks did not necessarily decrease over the past few years, but for some vulnerabilities remained constant.
Finally, we also demonstrate the practicality of our framework via an in-depth analysis on the recent Uniswap and Lendf.me attacks.

\end{abstract}
\setlength{\belowcaptionskip}{-10pt}
\section{Introduction}

As of today, Ethereum~\cite{wood2014ethereum} revolutionized the way digital assets are traded by being the first to introduce the concept of Turing-complete smart contracts on the blockchain.
These are programs that are executed and stored across the blockchain.
However, due to the tamper-resistant nature of blockchains, smart contracts can no longer be modified once deployed. 
At the time of writing, Ethereum has a market capitalization of over 42 billion USD, making it the second most valuable cryptocurrency on the market~\cite{ethereumMarket}.
As of writing, WETH, the most valuable Ethereum smart contract holds more than 2 billion USD worth of ether (Ethereum's own cryptocurrency)~\cite{wrappedEther}.
Moreover, Ethereum grew in the past 4 years from a daily transaction average of 10K in January 2016 to an average of 500K in January 2020~\cite{etherscanTransactions}.
Such an increase in value and popularity attracts abuse and the lack of a governing authority has led to a ``Wild West''-like situation, where several attackers began to exploit vulnerable smart contracts to steal their funds.
In the past, several smart contracts hosting tens of millions of USD were victims to attacks 
(\eg\cite{daohack,firstparity,secondparity}). 
Hence, over the past few years a rich corpus of research works and tools have surfaced to identify smart contract vulnerabilities (\eg\cite{Luu2016,tsankov2018securify,torres2018osiris,mueller2018,tikhomirov2018smartcheck,brent2018vandal,frank2020ethbmc,jiang2018contractfuzzer,kalra2018zeus}). However, most of these tools only focus on analyzing the bytecode of smart contracts and not their transactions or activities. Only a small number leverages transactions to detect attacks (\eg\cite{rodler2018sereum,chen2020soda,wu2020ethscope}), whereas the majority either requires the Ethereum client to be modified or large and complex attack detection scripts to be written.
Moreover, none of these tools allow to directly trace stolen assets after their detection.
\\
\indent
In this work, we introduce \textsc{Horus}, a framework capable of automatically detecting and analyzing smart contract attacks from historical blockchain data. 
Besides detecting attacks, the framework also provides means to quantify and trace the flow of stolen assets across Ethereum accounts.
The framework replays transactions without modifying the Ethereum client and encodes their execution as logical facts. Attacks are then detected using Datalog queries, making the framework easily extendable to detect new attacks.
Stolen funds are traced by loading detected transactions into a graph database and performing transaction graph analysis.
Using our framework, we conduct a longitudinal study that spans the entire past Ethereum blockchain history, from August 2015 to May 2020, consisting of over 3 million smart contracts.
One of the fundamental research questions we are investigating is whether these years of efforts have yielded visibly fewer attacks in the wild. 
If the tools proposed herein are effective, one could argue that attacks should have declined over time. To quantify the answer to this question, we start by investigating 
whether attacks occur continuously, or if they appear sporadically. While most well-known attacks carry significant monetary value, we wonder whether smaller, but ongoing attacks may occur more often and remain rather occluded. 
\\
\indent
\textbf{Contributions.}
We present the design and implementation of \textsc{Horus}, a framework that helps identifying smart contract attacks based on a sequence of blockchain transactions using Datalog queries. In addition, the framework extracts the quantity of stolen funds, including ether as well as tokens, and traces them across accounts to support behavioral studies of attackers.
We provide a longitudinal study on the security of Ethereum smart contracts of the past 4.5 years, and find 8,095 attacks in the wild, targeting a total of 1,888 vulnerable contracts.
Finally, we perform a forensic analysis of the recent Uniswap and Lendf.me hacks. 
\\
\indent
The remainder of the paper is organized as follows. Section~\ref{sec:background} introduces background on smart contracts and the Ethereum virtual machine. 
Section~\ref{sec:framework} presents our framework. 
Our evaluation is discussed in
Section~\ref{sec:evaluation}.
Section~\ref{sec:analysis} analyzes our results and presents our forensic analysis on the Uniswap and Lendf.me incidents.
Finally, Section~\ref{sec:relatedwork} and Section~\ref{sec:conclusion} discuss related work and conclude our paper, respectively.
\section{Background}
\label{sec:background}

\hspace{0.5cm}\textbf{Smart Contracts.}
Although, the notion of smart contracts is not new \cite{szabo1997}, the concept only became wide-spread with the release of Ethereum in 2015.
Ethereum smart contracts are fully-fledged programs that are different from traditional programs in  several ways.  
They are deterministic as they must be executed across a network of mutually distrusting nodes.
Once deployed, smart contracts cannot be removed or updated, unless they have been explicitly designed to do so. 
Furthermore, every smart contract has a balance that keeps track of the amount of ether owned by the contract, and a value storage that allows to keep state across executions.
They are usually developed using a high-level programming language, such as Solidity \cite{solidity}, that compiles into low-level bytecode. 
This bytecode is interpreted by the Ethereum Virtual Machine (EVM).

\textbf{Transactions.}
The deployment and execution of smart contracts occurs via transactions.
Smart contracts are identifiable via a unique 160-bit address that is generated during deployment.
Transactions may only be initiated by externally owned accounts\footnote{Externally owned accounts are accounts controlled via private keys that have no associated code.}.
Smart contract functions are triggered by encoding the function signature and arguments in the data field of a transaction.
A fallback function is executed whenever the provided function name is not implemented.
Transactions may also contain a given amount of ether that shall be transferred from one account to another.
Smart contracts may call other smart contracts during execution, thus, a single transaction may trigger further transactions, so-called internal transactions. 

\textbf{Ethereum Virtual Machine.}
The EVM is a stack-based virtual machine that supports a Turing-complete set of instructions allowing smart contracts to store data and interact with the blockchain.
The EVM uses a gas mechanism to associate costs to the execution of instructions. This guarantees termination and prevents denial-of-service attacks.
The EVM holds a machine state $\mu = (g,pc,m,i,s)$ during execution, where $g$ is the gas available, $pc$ is the program counter, $m$ represents the memory contents, $i$ is the active number of words in memory, and $s$ is the content of the stack.

\section{The \textsc{Horus} Framework}
\label{sec:framework}

\begin{figure}
    \centering
  	\includegraphics[scale=0.45]{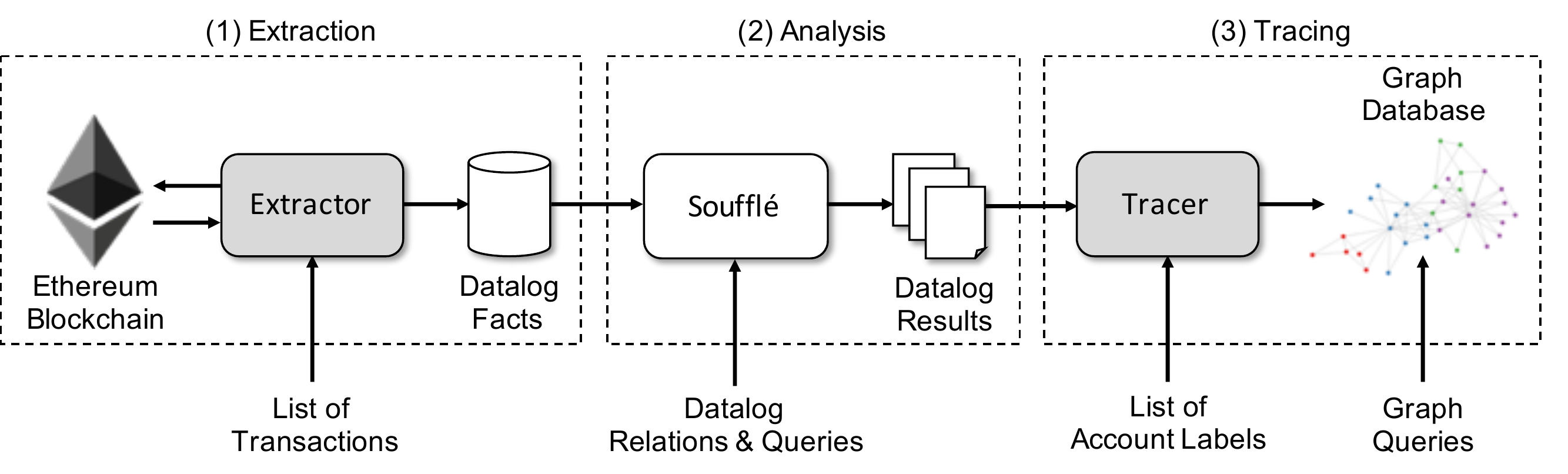}
	\caption{Architecture of \textsc{Horus}. Shaded boxes represent custom components, whereas boxes highlighted in white represent off-the-shelf components.}
	\label{fig:architecture}
\end{figure}

In this section, we provide details on the design and implementation of the \textsc{Horus} framework.
\textsc{Horus} automates the process of conducting longitudinal studies of attacks on Ethereum smart contracts.
The framework has the capability to detect and analyze smart contract attacks from historical data.
Moreover, the framework also provides means to trace the flow of stolen assets across Ethereum accounts.
The latter is particularly useful for studying the behavior of attackers.
\figname\ref{fig:architecture} provides an overview on the architecture of \textsc{Horus}.
The framework is organized as an EAT (extract, analyse, and trace) pipeline consisting of three different stages: 
\begin{itemize}
    \item[\textbf{(1)}] \textbf{Extraction:} The extraction stage takes as input a list of transactions from which execution related information is extracted and stored as Datalog facts.
    \item[\textbf{(2)}] \textbf{Analysis:} The analysis stage takes as input a set of Datalog relations and queries, which together identify attacks on the extracted Datalog facts. %
    \item[\textbf{(3)}] \textbf{Tracing:} The tracing stage retrieves a list of attacker accounts obtained via the analysis and fetches all transactions related to these accounts (including normal transactions, internal transactions and token transfers). Afterwards, a graph database is created, which captures the flow of funds (both ether and tokens) from and to these accounts. Further, the database can be augmented with a list of labeled accounts to enhance the tracing of stolen assets.
\end{itemize}

\noindent
In the following, we describe each of the three pipeline stages in more detail.
The entire framework was written in Python using roughly 2,000 lines of code\footnote{Code and data are publicly available at https://github.com/christoftorres/Horus.}. 

\subsection{Extraction}

The role of the extractor is to request from the Ethereum client the execution trace for a list of transactions and to convert them into logic relations that reflect the semantics of their execution.
An execution trace consists of an ordered list of executed EVM instructions.
Each record in that list contains information such as the executed opcode, program counter, call stack depth, and current stack values.
Unfortunately, execution traces cannot be obtained directly from historical blockchain data, they can only be recorded during contract execution.
Fortunately, the Go based Ethereum client (Geth) provides a debug functionality via the \texttt{debug\_traceTransaction} and \texttt{debug\_traceBlockByNumber} functions, which gives us the ability to replay the execution of any given past transaction or block and retrieve its execution trace. 
Execution traces are requested via Remote Procedure Call (RPC).
Previous works~\cite{rodler2018sereum,perez2019smart,chen2020soda,zhang2020txspector,wu2020ethscope}, did not rely on RPC as it is too slow. 
Instead, they modified Geth to speed up the process of retrieving execution traces. 
However, this has the limitation that users cannot use Geth's default version, but are required to use a modified version, and changes will need to be carried over every time a new version of Geth is released.
Moreover, at the time of writing, none of these works publicly disclosed their modified version of Geth, which not only makes it difficult to reproduce their results, but also to conduct future studies.
Therefore, rather than modifying Geth, we decided to improve the speed on the retrieval of execution traces via RPC. 
We noticed that execution traces contain a number of information that is irrelevant for our analysis. 
Fortunately, Geth allows us to inject our own execution tracer written in JavaScript~\cite{gethjavascript}.
Through this mechanism, we are able to reduce the size of the execution traces and improve execution speed, without actually modifying Geth.
For example, our JavaScript code removes the current program counter, the remaining gas and the instruction's gas cost from the execution trace.
Moreover, instead of returning a complete snapshot of the entire stack and memory for every executed instruction, our code only returns stack elements and memory slices that are relevant to the executed instruction.

\begin{lstlisting}[frame=single,language=DatalogFacts,caption={List of Datalog facts extracted by \textsc{Horus}.},captionpos=b,label={lst:datalog_facts}]
.decl opcode(step:number, op:Opcode, tx_hash:symbol)
.decl data_flow(step1:number, step2:number, tx_hash:symbol)
.decl arithmetic(step:number, op:Opcode, operand1:Value, operand2:Value, arithmetic_result:Value, evm_result:Value)
.decl storage(step:number, op:Opcode, tx_hash:symbol, caller:Address, contract:Address, index:Value, value:Value, depth:number)
.decl condition(step:number, tx_hash:symbol)
.decl erc20_transfer(step:number, tx_hash:symbol, contract:Address, from:Address, to:Address, value:Value)
.decl call(step:number, tx_hash:symbol, op:Opcode, caller:Address, callee:Address, input:symbol, value:Value, depth:number, call_id:number, call_branch:number, result:number)
.decl selfdestruct(step:number, tx_hash:symbol, caller:Address, contract:Address, destination:Address, value:Value)
.decl block(block_number:number, gas_used:number, gas_limit:number, timestamp:number)
.decl transaction(tx_hash:symbol, tx_index:number, block_number:number, from:Address, to:Address, input:symbol, gas_used:number, gas_limit:number, status:number)
\end{lstlisting}

\begin{figure}
    \centering
  	\includegraphics[width=0.65\linewidth]{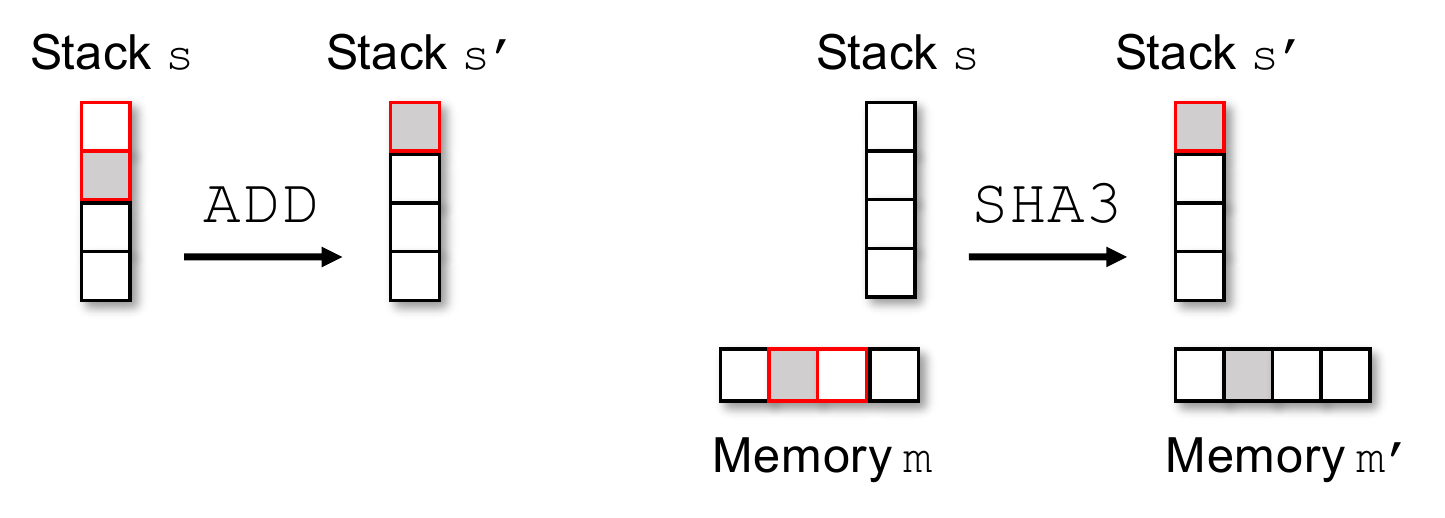}
	\caption{The example on the left depicts the propagation of taint via the \texttt{ADD} instruction, where the result pushed onto stack $s'$ becomes tainted because the second operand on stack $s$ was tainted. The example on the right depicts the propagation of taint via the \texttt{SHA3} instruction, where the result pushed onto stack $s'$ becomes tainted because the memory $m$ was tainted.}
	\label{fig:taintanalysis}
\end{figure}

Listing~\ref{lst:datalog_facts} shows the list of Datalog facts that our extractor produces by iterating through each of the records of the execution traces and encoding relevant information.
While most facts are related to low level EVM operations (\eg \textbf{\color{blue}{\texttt{call}}}), others are related to high level operations. 
For example, the \textbf{\color{blue}{\texttt{erc20\_transfer}}} fact refers to the ERC-20 token event ``Transfer'' that is emitted whenever tokens are transferred, where \textbf{\texttt{contract}} denotes the address of the token contract, and \textbf{\texttt{from}} and \textbf{\texttt{to}}, denote the sender and receiver of the tokens, respectively.
It is important to note that this list can easily be modified or extended to support different studies from the one proposed in this paper by modifying the extractor, analyzer and tracer.
Besides using the default types \textcolor{violet}{\texttt{number}} and \textcolor{violet}{\texttt{symbol}}, we also define our own three new types: \textcolor{teal}{\texttt{Address}} for 160-bit values, \textcolor{teal}{\texttt{Opcode}} for the set of EVM opcodes, and \textcolor{teal}{\texttt{Value}} for 256-bit stack values.
\\
\indent
\textbf{Dynamic Taint Analysis.}
The extractor leverages dynamic taint analysis to track the flow of data across instructions.
Security experts can then use the \textbf{\color{blue}{\texttt{data\_flow}}} fact to check if data flows from one instruction to another. 
Taint is introduced via sources, then propagated across the execution and finally checked if it flows into sinks.
Sources represent instructions that might introduce untrusted data (\eg\texttt{CALLDATALOAD} or \texttt{CALLDATACOPY}), whereas sinks represent instructions that are sensitive locations (\eg\texttt{CALL} or \texttt{SSTORE}).
We implemented our own dynamic taint analysis engine.
The engine loops through every executed instruction and checks whether the executed instruction is a source, for which the engine then introduces taint by tagging the affected stack value, memory region or storage location according to the semantics defined in~\cite{wood2014ethereum}.
We implemented the stack using an array structure following LIFO logic. 
Memory and storage are implemented using a Python dictionary that maps memory and storage addresses to values.
Taint propagation is performed at the byte level (see examples in \figname\ref{fig:taintanalysis}).

\textbf{Execution Order.}
Attacks such as the Parity wallets hacks were composed of two transactions being executed in a specific order. 
To detect such multi-transactional attacks, our framework encodes a total order across multiple transactions via the triplet $o = (b, t, s)$, where $b$ is the block number, $t$ is the transaction index, and $s$ is the execution step. 
The execution step is a simple counter that is reset at the beginning of the execution of a transaction and its value is incremented after each executed instruction.
An execution step is bound to a transaction index, which is on the other hand bound to a block number.
As such, our framework is able to precisely identify the execution order of any instruction across multiple transactions and the entire blockchain history.

\subsection{Analysis}

The second stage of our pipeline uses a Datalog engine to analyze whether a given list of Datalog relations and queries match any of the previously extracted Datalog facts.
These Datalog queries identify adversarial transactions, \ie malicious transactions that successfully carried out a concrete attack against a smart contract by exploiting a given vulnerability.
Our framework uses Soufflé as its Datalog engine.
Soufflé compiles Datalog relations and queries into a highly optimized C++ executable~\cite{jordan2016souffle}.
In the following, we provide Datalog queries for detecting reentrancy, Parity wallet hacks, integer overflows, unhandled exceptions and short address attacks.
Although, a number of smart contract vulnerabilities exist~\cite{atzei2017survey}, in this work we focus on those that are ranked by the NCC Group as the top 10 smart contract vulnerabilities~\cite{dasp} and for which we can extract the amount of ether or tokens that were either stolen or locked.
\\
\indent
\textbf{Reentrancy.}
Reentrancy occurs whenever a contract calls another contract, and the called contract calls back the original contract (\ie a re-entrant call) before the state in the original contract has been updated appropriately.
We detect reentrancy by identifying cyclic calls originating from the same caller and calling the same callee (see Listing~\ref{lst:reentrancy}).
We check if two successful \textcolor{blue}{\texttt{call}}s  (\ie result is \texttt{1}), share the same transaction \textcolor{violet}{\texttt{hash}}, \textcolor{violet}{\texttt{caller}}, \textcolor{violet}{\texttt{callee}}, \textcolor{violet}{\texttt{id}} and \textcolor{violet}{\texttt{branch}}, where the second call has a higher call \textcolor{violet}{\texttt{depth}} than the first call.
Afterwards, we check if there are two \textcolor{blue}{\texttt{storage}} operations with the same call depth as the first call, where the first operation is an \texttt{SLOAD} and occurs before the first call, and the second operation is an \texttt{SSTORE} and occurs after the second call. 

\begin{lstlisting}[frame=single,language=DatalogQuery,caption={Datalog query for detecting reentrancy attacks.},captionpos=b,label={lst:reentrancy}]
Reentrancy(hash, caller, callee, depth2, amount) :-
  storage(step1, "SLOAD", hash, _, caller, index, _, depth1),
  call(step2, hash, _, caller, callee, _,      _, depth1, id, branch, 1),
  call(step3, hash, _, caller, callee, _, amount, depth2, id, branch, 1),
  storage(step4, "SSTORE", hash, _, caller, index, _, depth1),
  depth1 < depth2, step1 < step2, step3 < step4, !match("0", amount).
\end{lstlisting}

\textbf{Parity Wallet Hacks.}
In this paper, we focus on detecting the two Parity wallet hacks~\cite{firstparity,secondparity}.
Both hacks were due faulty access control implementations that allowed attackers to set themselves as owners, which allowed them to perform critical actions such as the transfer of funds or the destruction of contracts.
We detect the first Parity wallet hack by checking if there exist two \textcolor{blue}{\texttt{transaction}}s $t_1$ and $t_2$, both containing the same sender and receiver, where the first 4 bytes of $t_1$'s input match the function signature of the \texttt{initWallet} function (\ie \texttt{e46dcfeb}), and 
if the first 4 bytes of $t_2$'s input match the function signature of the \texttt{execute} function (\ie \texttt{b61d27f6}) (see Listing~\ref{lst:parity_hack_1}).
Afterwards, we check whether there is a \textcolor{blue}{\texttt{call}}, which is part of $t_2$ and where $t_2$ is executed after $t_1$ (\ie \textcolor{violet}{\texttt{block1}} $<$ \textcolor{violet}{\texttt{block2}}; \textcolor{violet}{\texttt{block1}} $=$ \textcolor{violet}{\texttt{block2}}, \textcolor{violet}{\texttt{index1}} $<$ \textcolor{violet}{\texttt{index2}}). 
\begin{lstlisting}[frame=single,language=DatalogQuery,caption={Datalog query for detecting the first Parity wallet hack.},captionpos=b,label={lst:parity_hack_1}]
ParityWalletHack1(hash1, hash2, caller, callee, amount) :-
  transaction(hash1, index1, block1, from, to, input1, _, _, 1),
  substr(input1, 0, 8) = "e46dcfeb",
  transaction(hash2, index2, block2, from, to, input2, _, _, 1),
  substr(input2, 0, 8) = "b61d27f6",
  call(_, hash2, "CALL", caller, callee, _, amount, _, 1),
  (block1 < block2; block1 = block2, index1 < index2).
\end{lstlisting}
We detect the second Parity wallet hack in a very similar way to the first one, except that in this case we check if $t_2$'s input matches the function signature of the \texttt{kill} function (\ie \texttt{cbf0b0c0}) and  $t_2$ contains a \textcolor{blue}{\texttt{selfdestruct}} (see Listing~\ref{lst:parity_hack_2}).
\begin{lstlisting}[frame=single,language=DatalogQuery,caption={Datalog query for detecting the second Parity wallet hack.},captionpos=b,label={lst:parity_hack_2}]
ParityWalletHack2(hash1, hash2, contract, destination, amount) :-
  transaction(hash1, index1, block1, from, to, input1, _, _, 1),
  substr(input1, 0, 8) = "e46dcfeb",
  transaction(hash2, index2, block2, from, to, input2, _, _, 1),
  substr(input2, 0, 8) = "cbf0b0c0",
  selfdestruct(_, hash2, _, contract, destination, amount),
  (block1 < block2; block1 = block2, index1 < index2).
\end{lstlisting}

\textbf{Integer Overflows.}
We detect integer overflows by checking if data from \texttt{CALLDATALOAD} or \texttt{CALLDATACOPY} \textcolor{blue}{\texttt{opcode}}s flows into an \textcolor{blue}{\texttt{arithmetic}} operation, where the arithmetic result does not match the result returned by the EVM. Afterwards, we check whether the result of the arithmetic operation flows into an \texttt{SSTORE} \textcolor{blue}{\texttt{storage}} operation and an \textcolor{blue}{\texttt{erc20\_transfer}} occurs, where the \textcolor{violet}{\texttt{amount}} is one of the two operands used in the arithmetic computation (see Listing~\ref{lst:integer_overflow}). 
Please note that in this work, we only focus on detecting integer overflows related to ERC-20 tokens, since token smart contracts have been identified in the past to be frequent victims of integer overflows~\cite{bectoken,smttoken}.

\begin{lstlisting}[frame=single,language=DatalogQuery,caption={Datalog query for detecting integer overflow attacks.},captionpos=b,label={lst:integer_overflow}]
IntegerOverflow(hash, from, to, amount) :-
  (opcode(step1, "CALLDATALOAD", hash); 
   opcode(step1, "CALLDATACOPY", hash)),
  arithmetic(step2, _, operand1, operand2, arithmetic_res, evm_res),
  arithmetic_res != evm_res, (operand1 = amount; operand2 = amount),
  storage(step3, "SSTORE", hash, _, _, _, _, 1),
  data_flow(step1, step2, hash), data_flow(step2, step3, hash),
  erc20_transfer(_, hash, _, from, to, amount), !match("0", amount).
\end{lstlisting}

\textbf{Unhandled Exception.}
Inner calls executed by smart contracts may fail and by default only the state changes caused by those failed calls are rolled back.
It is the responsibility of the developer to check the result of every call and perform proper exception handling.
However, many developers forget or decide to ignore the handling of such exceptions, resulting in funds not being transferred to their rightful owners.
We detect an unhandled exception by checking whether a {\color{blue}\texttt{call}} with opcode \texttt{"CALL"} failed (\ie result is \texttt{0}) with an {\color{violet} \texttt{amount}} larger than zero and where the result was not used in a \textcolor{blue}{\texttt{condition}} (see Listing~\ref{lst:unhandled_exception}).

\begin{lstlisting}[frame=single,language=DatalogQuery,caption={Datalog query for detecting unhandled exceptions.},captionpos=b,label={lst:unhandled_exception}] 
UnhandledException(hash, caller, callee, amount) :-
  call(step, hash, "CALL", caller, callee, _, amount, _, 0),
  !match("0", amount), !used_in_condition(step, hash).
\end{lstlisting}

\textbf{Short Address.}
The ERC-20 functions \texttt{transfer} and \texttt{transferFrom} take as input a destination address and a given amount of tokens. During execution the EVM will add trailing zeros to the end of the transaction input if the transaction arguments are not correctly encoded as chunks of 32 bytes, thereby shifting the input bytes to the left by a few zeros, and therefore unwillingly increase the number of tokens to be transferred. However, attackers can exploit this fact by generating addresses that end with trailing zeros and omit these zeros, to then trick another party (\eg web service) into making a call to \texttt{transfer}/\texttt{transferFrom} containing the attacker's malformed address.
We detect a short address attack by first checking if the first 4 bytes of a \textcolor{blue}{\texttt{transaction}}'s input match either the function signature of \texttt{transfer} (\ie \texttt{a9059cbb}) or \texttt{transferFrom} (\ie \texttt{23b872dd}). Then, for the function \texttt{transfer} we check whether the length of the input is smaller than 68 (\ie 4 bytes function signature, 32 bytes destination address, and 32 bytes amount), and for the function \texttt{transferFrom} we check whether the length of the input is smaller than 100 (\ie 4 bytes function signature, 32 bytes from address, 32 bytes destination address, and 32 bytes amount), and finally we check if an \textcolor{blue}{\texttt{erc20\_transfer}} occurred (see Listing~\ref{lst:short_address}).

\begin{lstlisting}[frame=single,language=DatalogQuery,caption={Datalog query for detecting short address attacks.},captionpos=b,label={lst:short_address}] 
ShortAddress(hash, from, to, amount) :-
  transaction(hash, _, _, input, _, _, 1, _),
  (substr(input, 0, 8) = "a9059cbb", strlen(input) / 2 < 68;
   substr(input, 0, 8) = "23b872dd", strlen(input) / 2 < 100), 
  erc20_transfer(_, hash, _, from, to, amount), !match("0", amount).
\end{lstlisting}

\subsection{Tracing}

The final stage of our pipeline is the tracing of stolen assets, such as ether and tokens, from attacker accounts to labeled accounts (\eg exchanges). 
The tracer starts by extracting sender addresses and timestamps from malicious transactions that have been identified via the Datalog analysis.
Sender addresses are assumed to be accounts belonging to attackers.
Afterwards, the tracer uses Etherscan's API to retrieve for each sender address all its normal transactions, internal transactions and token transfers, and loads them into a Neo4j graph database.
We rely on a third-party service such as Etherscan to retrieve normal transactions, internal transactions and token transfers, because a default Ethereum node does not provide this functionality out-of-the-box.
Accounts are encoded as vertices and transactions as directed edges between those vertices.
We differentiate between three types of accounts: attacker accounts, unlabeled accounts, and labeled accounts.
Every account type contains an address.
Labeled accounts contain a category (\eg exchange) and a label (\eg Kraken 1).
We obtain categories and labels from Etherscan's large collection of labeled accounts\footnote{\url{https://etherscan.io/labelcloud}}.
We downloaded a total of 5,437 labels belonging to 204 categories.
We differentiate between three different types of transactions: normal transactions, internal transactions, and token transactions.
Each transaction type contains a transaction value, transaction hash, and transaction date. 
Token transactions contain a token name, token symbol and number of decimals. 
Transactions can be loaded either backwards or forwards.
Loading transactions forwards allows us to track where attackers sent their stolen funds to, whereas loading transactions backwards allows us to track where attackers received their funds from.
We start with the attacker's account when loading transactions and recursively load transactions for neighboring accounts that are part of the same transaction for up to a given number of hops.
We do not load transactions for accounts with more than 1,000 transactions.
This is to avoid bloating the graph database with transactions from mixing services, exchanges or gambling smart contracts. 
Moreover, when loading transactions backwards, we only load transactions that occurred before the timestamp of the attack, whereas when loading transactions forwards, we only load transactions that occurred after the timestamp of the attack.
Finally, when all transactions are loaded, security experts can query the graph database using Neo4j's own graph query language called Cypher, to trace the flow of stolen funds.
Evidently, our tracing is only effective up to a certain point, since mixing services and exchanges prevent further tracing. Nonetheless, our tracing is still useful to study whether attackers send their funds to mixers or exchanges and to identify which services are being used and to what extend.

\section{Evaluation}\label{sec:evaluation}

In this section, we demonstrate the scalability and effectiveness of our framework by performing a large-scale analysis of the Ethereum blockchain and comparing our results to those presented in previous works.%

\textbf{Dataset.}
We used the Ethereum ETL framework~\cite{ethereumetl} to retrieve a list of transactions for every smart contract deployed up to block 10 M.
We collected a total of 697,373,206 transactions and 3,362,876 contracts.
The deployment timestamps of the collected contracts range from August 7, 2015, to May 4, 2020.
We filtered out contracts without transactions and removed transactions that have a gas limit of 21,000 (\ie do not execute code). 
Moreover, similar to~\cite{rodler2018sereum}, we skipped all the transactions that were part of the 2016 denial-of-service attacks, as these incur high execution times~\cite{ethereumDoS2016}.
After applying these filters, we ended up with a final dataset of 1,234,197 smart contracts consisting of 371,419,070 transactions. 
During the extraction phase, \textsc{Horus} generated roughly 700GB of Datalog facts on the final dataset.

\textbf{Experimental Setup.}
All experiments were conducted using a machine with 64 GB of memory and an Intel(R) Core(TM) i7-8700 CPU with 12 cores clocked at 3.2 GHz, running 64-bit Ubuntu 18.04.5 LTS. 
Moreover, we used Geth version 1.9.9, Soufflé version 1.7.1, and Neo4j version 4.0.3.

\subsection{Results}

\begin{table}
\setlength{\tabcolsep}{5.6pt}
  \caption{Summary of detected vulnerable contracts and adversarial transactions.}
  \label{tbl:results}
  \centering
\begin{tabular}{l | r r | r r c}
    \multicolumn{1}{c}{} & \multicolumn{2}{c}{\textbf{Results}} & \multicolumn{3}{c}{\textbf{Validation}} \\
    \toprule
    \textbf{Vulnerability} & \textbf{Contracts} & \textbf{Transactions} & \multicolumn{1}{c}{\textbf{TP}} & \multicolumn{1}{c}{\textbf{FP}} & \multicolumn{1}{c}{\textbf{$p$}} \\
    \midrule
    Reentrancy & 46 & 2,508 & 45 & 1 & 0.97 \\
    Parity Wallet Hacks & 600 & 1,852 & 600 & 0 & 1.00 \\
    \hspace{3mm} Parity Wallet Hack 1 & 596 & 1,632 & 596 & 0 & 1.00 \\
    \hspace{3mm} Parity Wallet Hack 2 & 238 & 238 & 238 & 0 & 1.00 \\
    Integer Overflow & 125 & 443 & 65 & 0 & 1.00 \\
    \hspace{3mm} Overflow (\emph{Addition}) & 37 & 139 & 25 & 0 & 1.00 \\
    \hspace{3mm} Overflow (\emph{Multiplication}) & 23 & 120 & 20 & 0 & 1.00 \\
    \hspace{3mm} Underflow (\emph{Subtraction}) & 104 & 352 & 68 & 0 & 1.00 \\
    Unhandled Exception & 1,068 & 3,100 & 100 & 0 & 1.00 \\
    Short Address & 55 & 275 & 5 & 0 & 1.00 \\
    \midrule
    Total Unique & 1,888 & 8,095 &  &   &  \\
    \bottomrule
\end{tabular}
\end{table}

Table~\ref{tbl:results} summarizes our results: we found 1,888 attacked contracts and 8,095 adversarial transactions.
From these contracts, 46 were attacked using reentrancy, 600 were attacked during the Parity wallet hacks, 125 were attacked via integer overflows, 1,068 suffered from unhandled exceptions, and 55 were victims of short address attacks.
For the Parity wallet hacks, we find that the majority was attacked during the first hack.
We also observe that most contracts that are vulnerable to integer overflows, were attacked via an integer underflow.

\subsection{Validation}

We confirm our framework's correctness, by comparing our findings to those reported by previous works for which results were publicly available.
Also, we solely compare our finding to works that similarly to \textsc{Horus}, focus on detecting attacks rather than vulnerable contracts.
In cases where the results were not publicly available, we manually inspected the source code and transactions of flagged contracts using Etherscan.
Table~\ref{tbl:results} summarizes the results of our validation in terms of true positives (TP), false positives (FP) and precision ($p$).
Overall our framework achieves a high precision of 99.54\%.

\textbf{Reentrancy.}
First, we compare our results to those of \textsc{Sereum}~\cite{rodler2018sereum}. The authors reported a total of 16 vulnerable contracts, where 14 are false positives. The true positives include the DAO~\cite{dao} and the DSEthToken~\cite{dsethtoken} contract, which \textsc{Horus} has also identified.
\textsc{Horus} has flagged none of the 14 false positives.
Next, we compare our results to \textsc{{\AE}GIS}~\cite{ferreira2019aegis,ferreira2020aegis}.
\textsc{Horus} successfully detected the 7 contracts that were reported by \textsc{{\AE}GIS}.
Then, we compare our results to \textsc{SODA}~\cite{chen2020soda}.
\textsc{Horus} identified 25 of the 26 contracts that were flagged as true positives by \textsc{SODA}.
We analyzed the remaining contract (\texttt{0x59abb8006b30d7357869760d21b4965475198d9d}) and found that it is not vulnerable to reentrancy, which is in line with what other previous works discovered~\cite{wu2020ethscope}.
For the 5 false positives reported by \textsc{SODA}, we detected 3 of them, where two (\texttt{0xd4cd7c881f5ceece4917d856ce73f510d7d0769e} and \texttt{0x72f60eca0
db6811274215694129661151f97982e}) are actual true positives and have been misclassified by \textsc{SODA}. The other one (known as HODLWallet~\cite{hodlwallet}) is indeed a false positive.
Afterwards, we compare our results with those of \textsc{EthScope}~\cite{wu2020ethscope}. 
\textsc{Horus} detected 45 out of the 46 true positives reported by \textsc{EthScope}. 
The non-reported contract is the DarkDAO~\cite{darkdao}, which did not suffer from a reentrancy attack and is, therefore, a false positive.
In terms of false positives, \textsc{Horus} only has one in common with \textsc{EthScope}, namely the aforementioned HODLWallet contract. The other two false positives that \textsc{EthScope} reported were correctly identified as true negatives by \textsc{Horus}.
Finally, we compare our results with those of Zhou et al.~\cite{zhou2020ever}.
\textsc{Horus} found 22 of the 26 contracts that have been reported as true positives by Zhou et al. 
We inspected the remaining 4 contracts and found that they are false positives.
\\
\indent
\textbf{Parity Wallet Hacks.} 
For the first Parity wallet hack, we compared our results to those reported by \textsc{{\AE}GIS} and Zhou et al. 
\textsc{{\AE}GIS} reported 3 contracts, which have also been found by \textsc{Horus}. 
Next, Zhou et al.~reported 622 contracts, of which \textsc{Horus} found 596. We analyzed the remaining 26 contracts and found that these are false positives. After analyzing their list of transactions, we could not find evidence of the two exploiting transactions, namely \texttt{initWallet} and \texttt{execute}.
For the second Parity wallet hack, we compared our results to those of \textsc{{\AE}GIS}. \textsc{Horus} found 238 contracts, of which 236 were also reported by \textsc{{\AE}GIS}. The remaining two are true positives and have not been identified by \textsc{{\AE}GIS}.
\\
\indent
\textbf{Integer Overflow.}
We compared our findings to those of Zhou et al. The authors found 50 contracts, whereas we found 125 contracts. \textsc{Horus} detected 49 of the 50 contracts reported by Zhou et al. We analyzed the undetected contract (\texttt{0xa9a8ec071ed0ed5be571396438a046a423a0c206}) and found no evidence of an integer overflow.
Besides our comparison with Zhou et al., we also tried to analyze manually the source code of the reported contracts. We were able to obtain the source code for 65 of the 125 reported contracts. Our manual inspection identified that all of the contracts are true positives. They either contained a faulty arithmetic check or no arithmetic check at all.
\\
\indent
\textbf{Unhandled Exception.}
Since none of the previous works analyzed unhandled exceptions, we manually analyzed the source code of the contracts reported by \textsc{Horus}. However, we limited our validation to a random sample of 100 contracts since manually analyzing 1,068 contracts is infeasible.
We find that all of the 100 contracts contained in their source code either a direct call or a function call that did not check the return value. Therefore, we conclude that \textsc{Horus} reports no false positives on the detection of unhandled exceptions.
\\
\indent
\textbf{Short Address.} We compared our results to those reported by \textsc{SODA}.
\textsc{SODA} detected 726 contracts and 6,599 transactions, whereas \textsc{Horus} detected 55 contracts and 275 transactions. 
After further investigation, we found that the contracts and transactions detected by 
\textsc{Horus} were also detected by \textsc{SODA}.
We also found that \textsc{SODA} reported transactions that failed or where the transferred amount was zero, while \textsc{Horus} only reported transactions that were successful and where an ERC-20 transfer event was successfully triggered with an amount larger than zero.
Moreover, we were able to obtain the source code for 5 of the reported contracts and confirm that the \texttt{transfer} or \texttt{transferFrom} functions contained inside those contracts do not validate the input length of parameters.

\section{Analysis}
\label{sec:analysis}

In this section, we demonstrate the practicality of \textsc{Horus} in detecting and analyzing real-world smart contract attacks via an analysis of our evaluated results and a case study on the recent Uniswap and Lendf.me incidents.

\subsection{Volume and Frequency of Attacks}

\vspace{-5mm} 
\begin{figure}
    \centering
  	\includegraphics[width=1.0\textwidth]{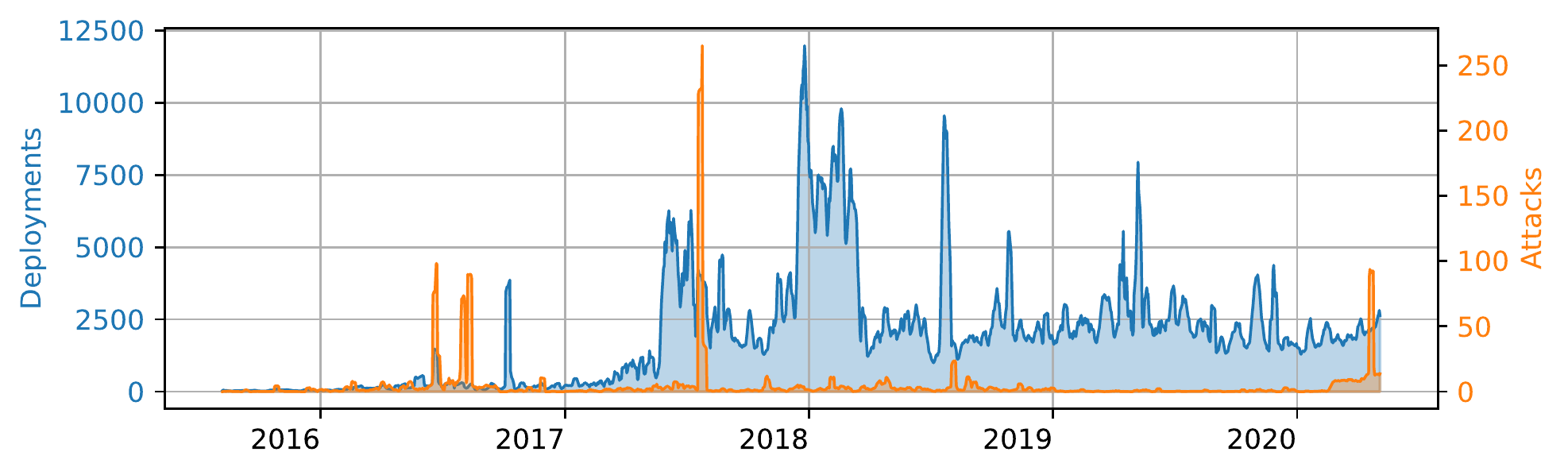}
	\caption{Weekly average of daily contract deployments and attacks over time.}
	\label{fig:deployemnts}
\end{figure}

\figname\ref{fig:deployemnts} depicts the weekly average of daily attacks in comparison to the weekly average of daily deployments.
We state that the peak of weekly deployed contracts was at the end of 2017, and that the largest volume of weekly attacks occurred before this peak.
Moreover, most attacks seem to occur in clusters of the same day. We suspect that attackers scan the blockchain for similar vulnerable contracts and exploit them at the same time.
The first three spikes in the attacks correspond to the DAO and Parity wallet hacks, whereas the last spike corresponds to the recent Uniswap/Lendf.me hacks.

\vspace{-5mm} 
\begin{figure}
    \centering
  	\includegraphics[width=1.0\textwidth]{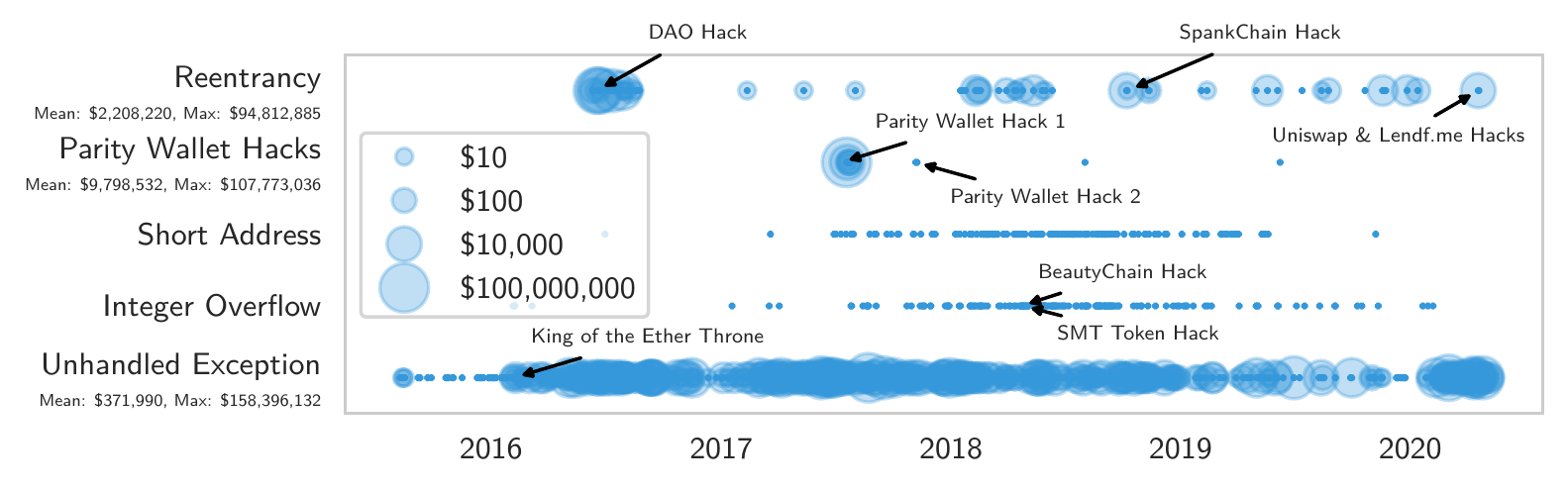}
	\caption{Volume and frequency of smart contract attacks over time.}
	\label{fig:timeline}
\end{figure}

\figname\ref{fig:timeline} depicts the occurrences of adversarial transactions per vulnerability type that we measured during our evaluation. While reentrancy attacks seem to occur more sporadically, other types of vulnerabilities such as unhandled exceptions are triggered rather continuously. 
Overall, we see that over time less contracts became victims to short address attacks and integer overflows, suggesting that smart contracts have become more secure over the past few years.
However, we also see that smart contracts still remain vulnerable to well-known vulnerabilities such as reentrancy and unhandled exceptions, despite automated security tools being available. 
\figname\ref{fig:timeline} also illustrates for each adversarial transaction the amount of USD that was either stolen (reentrancy and Parity wallet hack 1) or locked (unhandled exception and Parity wallet hack 2).
The USD amounts were calculated by multiplying the price of one ether at the time of the attack times the ether extracted via our Datalog query.
We do not provide USD amounts for short address attacks and integer overflows, because these attacks involve stolen ERC-20 tokens and we were not able to obtain the historical prices of these tokens.
We can see that the DAO hack and the first Parity wallet hack remain the two most devastating attacks in terms of ether stolen, with ether worth 94,812,885 USD and 107,773,036 USD, respectively.
We marked well-known incidents such as the DAO hack, or the two Parity wallet hacks for the reader's convenience and to demonstrate that \textsc{Horus} is able to detect them.

\subsection{Forensic Analysis on Uniswap and Lendf.me Incidents}
\label{sec:dex_incidents}

\vspace{-5mm} 
\begin{figure}
    \centering
  	\includegraphics[width=0.85\textwidth]{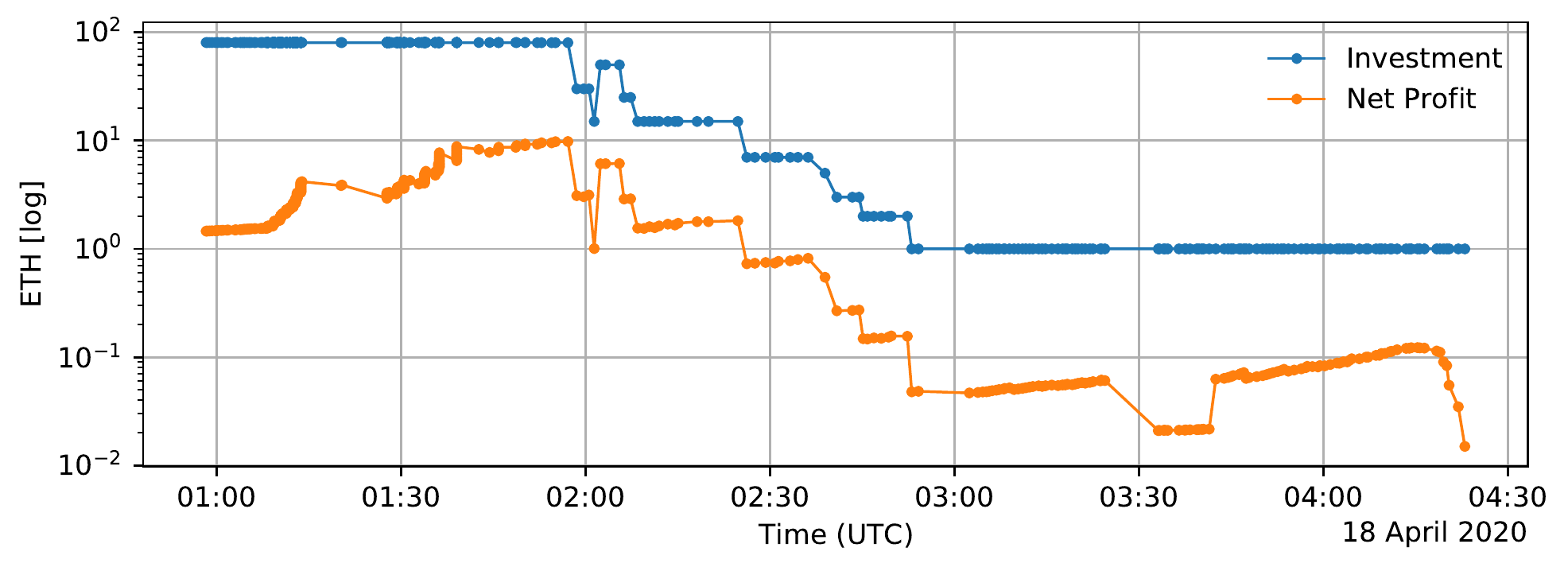}
	\caption{Invested ETH and net profit made by Uniswap attackers over time.}
	\label{fig:uniswap_profit}
\end{figure}

\indent
\textbf{Uniswap.}
On April 18, 2020, attackers were able to drain a large amount of ether from Uniswap's liquidity pool of ETH-imBTC~\cite{uniswap}.
They purposely chose the imBTC token as it implements the ERC777 standard, which would allow them to register a callback function and therefore perform a reentrancy attack on Uniswap.
The attackers would start by purchasing imBTC tokens for ETH.
Afterward, they would exchange half of the purchased imBTC tokens within the same transaction back to ETH.
However, the latter would trigger a callback function that the attackers registered before the attack, allowing them to take control and call back the Uniswap contract to exchange the remaining half of imBTC tokens to ETH before the conversion rate was updated. 
Thus, the attackers could trade the second batch of imBTC tokens at a more profitable conversion rate.
Interestingly, this vulnerability was known to Uniswap and was publicly disclosed precisely a year before the attack~\cite{uniswapConsensys}.
\\
\indent
We used \textsc{Horus} to extract and analyze all the transactions mined on that day, and identified a total of 525 transactions performing reentrancy attacks against Uniswap with an accumulated profit of 1,278 ETH (232,239.46 USD).
The attack began at 00:58:19 UTC and ended roughly 3.5 hours later at 04:22:58 UTC.
\figname\ref{fig:uniswap_profit} depicts a timeline of the attack, showing the amount of ether that the attackers invested and the net profit they made per transaction. 
We see that the net profit goes down over time.
The highest profit made for a single transaction was roughly 9.79 ETH (1,778.72 USD), while the lowest profit was 0.01 ETH (2.73 USD).
The attackers began their attack by purchasing tokens for roughly 80 ETH and went over time down to 1 ETH. 
Moreover, we see that the profit was mostly tied to the amount of ether that the attackers were investing (\ie using to purchase imBTC tokens). However, we also see that sometimes there were some fluctuations, where the attackers were making more profit while they would invest the same amount of ether. This is probably due to other participants trading imBTC on Uniswap during the attack and therefore influencing the exchange rates.
In the last step, we traced the entire ether flow from the attackers account for up to 5 hops using \textsc{Horus}'s tracing capabilities. 
Our transaction graph analysis reveals that the attackers exchanged roughly 702 ETH (55\% of the stolen funds) for tokens on different exchanges: 589 ETH on Uniswap for WETH, DAI, USDC, BAT, and MKR, 31 ETH on Compound, and 82 ETH on 1inch.exchange. The latter is of particular interest for law enforcement agencies as 1inch.exchange keeps track of IP addresses of transactions performed over their platform~\cite{1inchExchange}, which can be useful in deanonymizing the attackers.
\begin{figure}[H]
\centering
\begin{subfigure}[c]{0.53\linewidth}
    \centering
\includegraphics[width=1\linewidth]{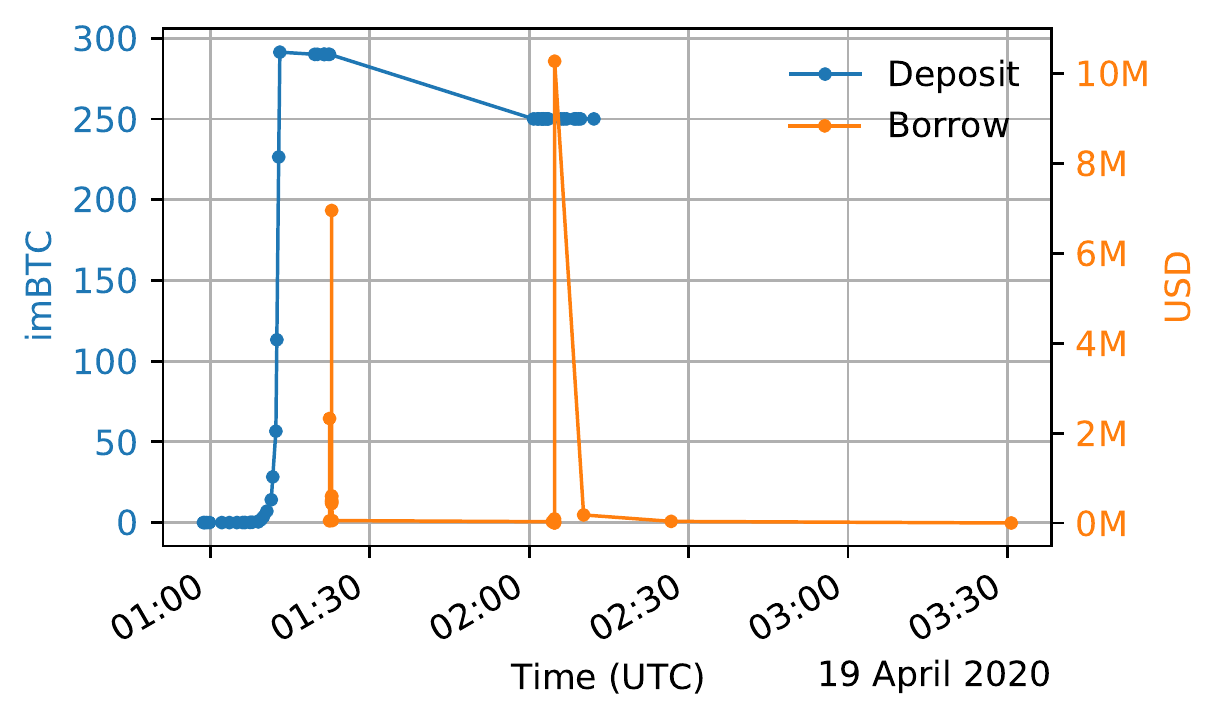}
\end{subfigure}
\begin{subfigure}[c]{0.44\linewidth}
    \centering
\includegraphics[width=1\linewidth]{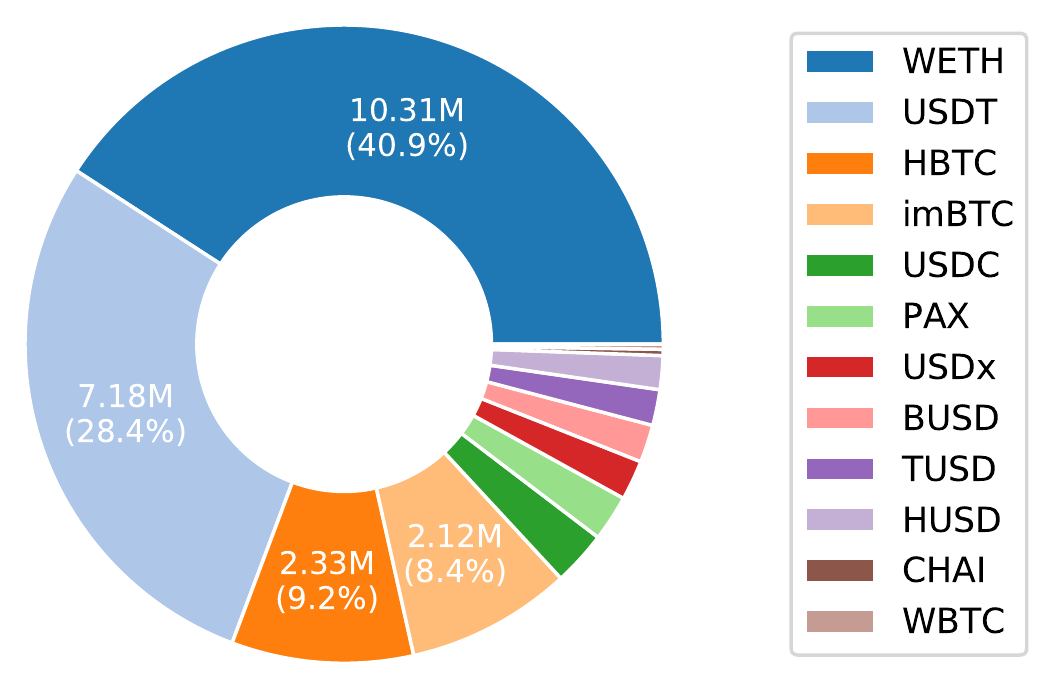}
\end{subfigure}
\caption{Deposited and borrowed tokens by Lendf.me attackers over time.}
\label{fig:lendfme_hack}
\end{figure}
\textbf{Lendf.me.} On April 19, 2020, attackers were able to drain all of Lendf.me's liquidity pools~\cite{lendfme}.
Similar to the Uniswap hack, the attackers exploited the fact that Lendf.me was trading imBTC and could register a callback function to perform a reentrancy attack. 
The attackers would start by depositing $x$ amount of imBTC tokens into Lendf.me's liquidity pool. 
Next, still within the same transaction, they would deposit another amount $y$, however, this time triggering the callback function registered by the attackers, which would withdraw the previously deposited $x$ tokens from Lendf.me.
By the end of the transaction, the imBTC balance of the attackers on the imBTC token contract would be $x-y$, but the imBTC balance on the Lendf.me contract would be $x+y$, thereby increasing their imBTC balance on Lendf.me by $x$ without actually depositing it.
Similar to Uniswap, the issue here is that the user’s balance is only updated after the transfer of tokens, thus the update is based on data before the transfer and therefore ignoring any updates made in between.
\\
\indent
Using \textsc{Horus}, we extracted and analyzed all the transactions mined on that day. We identified a total of 46 transactions performing reentrancy attacks against Lendf.me, and 19 transactions using the stolen imBTC tokens to borrow other tokens.
\figname\ref{fig:lendfme_hack} shows on the left the amount of imBTC tokens that the attackers deposited during the attack and the amount of USD that the attackers made by borrowing other tokens.
The right-hand side of \figname\ref{fig:lendfme_hack} depicts the number of tokens in USD that the attackers borrowed from Lendf.me. The attackers borrowed from 12 different tokens, worth together 25,244,120.74 USD, where 10.31M USD are only from borrowing WETH.
The attackers launched their attack at 00:58:43 UTC and stopped 2 hours later at 02:12:11 UTC.
They started depositing low amounts of imBTC and increased their amounts over time up to 291.35 imBTC.
The borrowing started at 01:22:27 UTC and ended at 03:30:42 UTC.
Finally, we used \textsc{Horus} to trace the flow of tokens from the attackers account for up to 3 hops.
We found that the attackers initially traded some parts of the stolen tokens for other tokens on ParaSwap, Compound, Aave, and 1inch.exchange.
However, at 14:16:52 UTC, thus about 10 hours later, the attackers started sending all the stolen tokens back to Lendf.me's admin account (\texttt{0xa6a6783828ab3e4a9db54302bc01c4ca73f17efb}). Lendf.me then moved all the tokens into a recovery account (\texttt{0xc88fcc12f400a0a2cebe87110dcde0dafd2
9f148}) where users could then reclaim their tokens.

\section{Related Work}\label{sec:relatedwork}

\indent \indent \textbf{Static Analysis.}
Researchers proposed a number of tools to detect smart contract vulnerabilities via static analysis.
Luu \etal\cite{Luu2016} proposed \textsc{Oyente}, the first symbolic execution tool for smart contracts.
Other tools such as \textsc{Osiris}~\cite{torres2018osiris}, combine symbolic execution and taint analysis to detect integer bugs.
\textsc{Mythril}\\\cite{mueller2018} uses a mix of symbolic execution and control-flow checking.
\textsc{Maian}~\cite{nikolic2018finding} employs inter-procedural symbolic execution.
\textsc{teEther}~\cite{krupp2018teether} automatically generates exploits for smart contracts.
\textsc{HoneyBadger}~\cite{torres2019art} performs symbolic execution to detect honeypots. 
However, symbolic execution is often unable to explore all program states, making it generally unsound.
Formal verification tools were proposed \cite{mavridou2018tool,wang2019formal}, together with a formal definition of the EVM \cite{hildenbrandt2017kevm}.
\textsc{EthBMC}~\cite{frank2020ethbmc} uses bounded model checking to detect vulnerabilities, whereas
\textsc{eThor}~\cite{schneidewind2020ethor} uses reachability analysis.
\textsc{Zeus}~\cite{kalra2018zeus} verifies the correctness of smart contracts using abstract interpretation and model checking.
\textsc{SmartCheck}~\cite{tikhomirov2018smartcheck} checks Solidity source code against XPath patterns.
\textsc{VeriSmart}~\cite{so2020verismart} leverages counter example-based inductive synthesis to detect arithmetic bugs.
\textsc{Securify}~\cite{tsankov2018securify} extracts semantic information from the dependency graph to check for compliance and violation patterns using Datalog.
\textsc{Vandal}~\cite{brent2018vandal} converts EVM bytecode to semantic logic relations and checks them against Datalog queries. 
The main difference between these works and ours, is that they analyze the bytecode of smart contracts, whereas we analyze the execution of transactions.
\\
\indent
\textbf{Dynamic Analysis.}
Although less apparent, a number of dynamic approaches have also been proposed.
\textsc{ECFChecker}~\cite{grossman2017online} enables the runtime detection of reentrancy attacks via a modified EVM.
\textsc{Sereum}~\cite{rodler2018sereum} proposes a modified EVM to protect deployed smart contracts against reentrancy attacks.
\textsc{{\AE}GIS}~\cite{ferreira2019aegis,ferreira2020aegis} presents a smart contract and a DSL to protect against all kinds of runtime attacks.
\textsc{SODA}~\cite{chen2020soda} uses a modified Ethereum client to inject custom modules for the online detection of malicious transactions. 
Perez \etal\cite{perez2019smart} use Datalog to study the transactions of vulnerable smart contracts that have been detected by previous works.
\textsc{EthScope}~\cite{wu2020ethscope} loads historical data into an Elasticsearch database and adds dynamic taint analysis to the client to analyze transactions.
Zhou et al.~\cite{zhou2020ever} study attacks and defenses by encoding transactional information as action trees and result graphs.
\textsc{TxSpector}~\cite{zhang2020txspector} is a concurrent work to ours and adopts the Datalog facts proposed by \textsc{Vandal}. 
However, these facts were designed to analyze bytecode and do not allow to detect multi-transactional attacks.
In contrast to these works, our work does not modify the Ethereum client. 
Instead, we dynamically inject our custom tracer into the client.
We also provide a new set of Datalog facts that allow to check for multi-transactional attacks and describe data flows between instructions via dynamic taint analysis.
Finally, none of the aforementioned tools provide means to trace stolen assets across the Ethereum blockchain.

\section{Conclusion}\label{sec:conclusion}

A wealth of automated vulnerability detection tools for Ethereum smart contracts were proposed over the past years. This raises the question whether the security of smart contracts has improved.
In this paper, we presented the design and implementation of an extensible framework for carrying out longitudinal studies on detecting, analyzing, and tracing of smart contract attacks.
We analyzed transactions from August 2015 to May 2020 and identified 8,095 attacks as well as 1,888 vulnerable contracts. Our analysis revealed that while the number of attacks seems to have decreased for attacks such as integer overflows, unhandled exceptions and reentrancy attacks still seem to remain present despite an abundance of new smart contract security tools.
Finally, we also presented an in-depth analysis on the recent Uniswap and Lendf.me incidents.

\section*{Acknowledgments}

We would like to thank the anonymous reviewers and Johannes Krupp for their valuable comments and feedback. This work was partly supported by the Luxembourg National Research Fund (FNR) under grant 13192291.

\bibliographystyle{splncs04}
\bibliography{references}

\begin{thebibliography}{10}
\providecommand{\url}[1]{\texttt{#1}}
\providecommand{\urlprefix}{URL }
\providecommand{\doi}[1]{https://doi.org/#1}

\bibitem{atzei2017survey}
Atzei, N., Bartoletti, M., Cimoli, T.: A survey of attacks on ethereum smart
  contracts (sok). In: International Conference on Principles of Security and
  Trust. pp. 164--186. Springer (2017)

\bibitem{brent2018vandal}
Brent, L., Jurisevic, A., Kong, M., Liu, E., Gauthier, F., Gramoli, V., Holz,
  R., Scholz, B.: Vandal: A scalable security analysis framework for smart
  contracts. arXiv preprint arXiv:1809.03981  (2018)

\bibitem{chen2020soda}
Chen, T., Cao, R., Li, T., Luo, X., Gu, G., Zhang, Y., Liao, Z., Zhu, H., Chen,
  G., He, Z., et~al.: Soda: A generic online detection framework for smart
  contracts. In: Proceedings of the Network and Distributed System Security
  Symposium ({NDSS'20}) (2020)

\bibitem{ethereumMarket}
CoinMarketCap: {Top 100 Cryptocurrencies by Market Capitalization} (September
  2020), https://coinmarketcap.com

\bibitem{uniswapConsensys}
{ConsenSys Diligence}: {Uniswap Audit} (September 2019),
  https://github.com/ConsenSys/Uniswap-audit-report-2018-12

\bibitem{dao}
{Etherscan}: {The DAO} (2016),
  https://etherscan.io/address/0xbb9bc244d798123fde 783fcc1c72d3bb8c189413

\bibitem{darkdao}
{Etherscan}: {The Dark DAO} (2016),
  https://etherscan.io/address/0x304a554a310c7e 546dfe434669c62820b7d83490

\bibitem{dsethtoken}
{Etherscan}: {DSEthToken} (2017),
  https://etherscan.io/address/0xd654bdd32fc99471 455e86c2e7f7d7b6437e9179

\bibitem{hodlwallet}
{Etherscan}: {HODLWallet} (2018),
  https://etherscan.io/address/0x4a8d3a662e0fd6a8 bd39ed0f91e4c1b729c81a38

\bibitem{lendfme}
{Etherscan}: {Lendf.Me - MoneyMarket} (2019),
  https://etherscan.io/address/0x0eee 3e3828a45f7601d5f54bf49bb01d1a9df5ea

\bibitem{uniswap}
{Etherscan}: {Uniswap: imBTC} (2019),
  https://etherscan.io/address/0xffcf45b540e6c 9f094ae656d2e34ad11cdfdb187

\bibitem{etherscanTransactions}
{Etherscan}: {Ethereum Daily Transactions Chart} (September 2020),
  https://etherscan.io/chart/tx

\bibitem{wrappedEther}
Etherscan: {Wrapped Ether} (September 2020),
  https://etherscan.io/address/0xc02aaa39b223fe8d0a0e5c4f27ead9083c756cc2

\bibitem{ferreira2019aegis}
Ferreira~Torres, C., Baden, M., Norvill, R., Jonker, H.: {{\AE}GIS: Smart
  Shielding of Smart Contracts}. In: Proceedings of the 2019 ACM SIGSAC
  Conference on Computer and Communications Security. pp. 2589--2591 (2019)

\bibitem{torres2018osiris}
Ferreira~Torres, C., Sch\"{u}tte, J., State, R.: Osiris: Hunting for integer
  bugs in ethereum smart contracts. In: Proceedings of the 34th Annual Computer
  Security Applications Conference. pp. 664--676. ACSAC '18, ACM, New York, NY,
  USA (2018). \doi{10.1145/3274694.3274737}

\bibitem{ferreira2020aegis}
Ferreira~Torres, C., Steichen, M., Norvill, R., Fiz~Pontiveros, B., Jonker, H.:
  {{\AE}GIS: Shielding Vulnerable Smart Contracts Against Attacks}. In:
  Proceedings of the 15th ACM Asia Conference on Computer and Communications
  Security (ASIA CCS’20), October 5--9, 2020, Taipei, Taiwan (2020)

\bibitem{torres2019art}
Ferreira~Torres, C., Steichen, M., State, R.: {The Art of The Scam:
  Demystifying Honeypots in Ethereum Smart Contracts}. In: 28th {USENIX}
  Security Symposium ({USENIX} Security 19). pp. 1591--1607. {USENIX}
  Association, Santa Clara, CA (Aug 2019)

\bibitem{frank2020ethbmc}
Frank, J., Aschermann, C., Holz, T.: {ETHBMC}: A bounded model checker for
  smart contracts. In: 29th {USENIX} Security Symposium ({USENIX} Security 20).
  {USENIX} Association, Boston, MA (Aug 2020)

\bibitem{grossman2017online}
Grossman, S., Abraham, I., Golan-Gueta, G., Michalevsky, Y., Rinetzky, N.,
  Sagiv, M., Zohar, Y.: Online detection of effectively callback free objects
  with applications to smart contracts. Proceedings of the ACM on Programming
  Languages  \textbf{2}(POPL), ~48 (2017)

\bibitem{hildenbrandt2017kevm}
{Hildenbrandt}, E., {Saxena}, M., {Rodrigues}, N., {Zhu}, X., {Daian}, P.,
  {Guth}, D., {Moore}, B., {Park}, D., {Zhang}, Y., {Stefanescu}, A., {Rosu},
  G.: Kevm: A complete formal semantics of the ethereum virtual machine. In:
  2018 IEEE 31st Computer Security Foundations Symposium (CSF). pp. 204--217
  (2018)

\bibitem{jiang2018contractfuzzer}
Jiang, B., Liu, Y., Chan, W.: Contractfuzzer: Fuzzing smart contracts for
  vulnerability detection. In: Proceedings of the 33rd ACM/IEEE International
  Conference on Automated Software Engineering. pp. 259--269. ACM (2018)

\bibitem{jordan2016souffle}
Jordan, H., Scholz, B., Suboti{\'c}, P.: Souffl{\'e}: On synthesis of program
  analyzers. In: International Conference on Computer Aided Verification. pp.
  422--430. Springer (2016)

\bibitem{kalra2018zeus}
Kalra, S., Goel, S., Dhawan, M., Sharma, S.: Zeus: Analyzing safety of smart
  contracts. In: NDSS. pp. 1--12 (2018)

\bibitem{daohack}
{Klint Finley}: {A \$50 Million Hack Just Showed That the DAO Was All Too
  Human} (June 2016),
  https://www.wired.com/2016/06/50-million-hack-just-showed-dao-human/

\bibitem{krupp2018teether}
Krupp, J., Rossow, C.: teether: Gnawing at ethereum to automatically exploit
  smart contracts. In: 27th {USENIX} Security Symposium ({USENIX} Security 18).
  pp. 1317--1333 (2018)

\bibitem{Luu2016}
Luu, L., Chu, D.H., Olickel, H., Saxena, P., Hobor, A.: {Making Smart Contracts
  Smarter}. In: Proceedings of the 2016 ACM SIGSAC Conference on Computer and
  Communications Security - CCS'16. pp. 254--269. ACM Press, New York, New
  York, USA (2016). \doi{10.1145/2976749.2978309}

\bibitem{mavridou2018tool}
Mavridou, A., Laszka, A.: Tool demonstration: Fsolidm for designing secure
  ethereum smart contracts. In: International Conference on Principles of
  Security and Trust. pp. 270--277. Springer (2018)

\bibitem{ethereumetl}
{Medvedev, Evgeny}: {Ethereum ETL v1.3.0} (April 2019),
  https://github.com/blockchain-etl/ethereum-etl

\bibitem{mueller2018}
Mueller, B.: Smashing ethereum smart contracts for fun and real profit.(2018).
  In: The 9th annual HITB Security Conference (2018)

\bibitem{dasp}
{NCC Group}: {Decentralized Application Security Project (or DASP) Top 10 of
  2018} (December 2018), https://dasp.co

\bibitem{nikolic2018finding}
Nikoli{\'c}, I., Kolluri, A., Sergey, I., Saxena, P., Hobor, A.: Finding the
  greedy, prodigal, and suicidal contracts at scale. In: Proceedings of the
  34th Annual Computer Security Applications Conference. pp. 653--663. ACM
  (2018)

\bibitem{bectoken}
{PeckShield - batchOverflow Bug}: {ALERT: New batchOverflow Bug in Multiple
  ERC20 Smart Contracts (CVE-2018-10299)} (April 2018),
  https://blog.peckshield.com/2018/04/22/batchOverflow/

\bibitem{smttoken}
{PeckShield - proxyOverflow Bug }: {New proxyOverflow Bug in Multiple ERC20
  Smart Contracts (CVE-2018-10376)} (April 2018),
  https://blog.peckshield.com/2018/04/25/proxyOverflow/

\bibitem{perez2019smart}
Perez, D., Livshits, B.: Smart contract vulnerabilities: Vulnerable does not
  imply exploited. In: 30th {USENIX} Security Symposium ({USENIX} Security 21).
  {USENIX} Association, Vancouver, B.C. (Aug 2021)

\bibitem{secondparity}
Petrov, S.: Another parity wallet hack explained (nov 2017),
  https://medium.com/@Pr0Ger/another-parity-wallet-hack-explained-847ca46a2e1c

\bibitem{rodler2018sereum}
Rodler, M., Li, W., Karame, G., Davi, L.: Sereum: Protecting existing smart
  contracts against re-entrancy attacks. In: Proceedings of the Network and
  Distributed System Security Symposium ({NDSS'19}) (2019)

\bibitem{1inchExchange}
{Ryan Sean Adams}: {} (September 2020), https://twitter.com/RyanSAdams/status/
  1252574107159408640

\bibitem{schneidewind2020ethor}
Schneidewind, C., Grishchenko, I., Scherer, M., Maffei, M.: ethor: Practical
  and provably sound static analysis of ethereum smart contracts. arXiv
  preprint arXiv:2005.06227  (2020)

\bibitem{so2020verismart}
So, S., Lee, M., Park, J., Lee, H., Oh, H.: Verismart: A highly precise safety
  verifier for ethereum smart contracts. In: 2020 IEEE Symposium on Security
  and Privacy (SP). pp. 1678--1694. IEEE (2020)

\bibitem{ethereumDoS2016}
StackExchange: Why is my node synchronization stuck/extremely slow at block
  2,306,843? (November 2016),
  https://ethereum.stackexchange.com/questions/9883/why-is-my-node-synchronization-stuck-extremely-slow-at-block-2-306-843

\bibitem{szabo1997}
Szabo, N.: Formalizing and securing relationships on public networks. First
  Monday  \textbf{2}(9) (1997)

\bibitem{gethjavascript}
Szilágyi, P.: {Go-Ethereum Management APIs - JavaScript-based tracing} (jan
  2020),
  https://github.com/ethereum/go-ethereum/wiki/Management-APIs\#javascript-based-tracing

\bibitem{tikhomirov2018smartcheck}
Tikhomirov, S., Voskresenskaya, E., Ivanitskiy, I., Takhaviev, R., Marchenko,
  E., Alexandrov, Y.: Smartcheck: Static analysis of ethereum smart contracts.
  In: 2018 IEEE/ACM 1st International Workshop on Emerging Trends in Software
  Engineering for Blockchain (WETSEB). pp. 9--16. IEEE (2018)

\bibitem{tsankov2018securify}
Tsankov, P., Dan, A., Drachsler-Cohen, D., Gervais, A., Buenzli, F., Vechev,
  M.: Securify: Practical security analysis of smart contracts. In: Proceedings
  of the 2018 ACM SIGSAC Conference on Computer and Communications Security.
  pp. 67--82. ACM (2018)

\bibitem{wang2019formal}
Wang, Y., Lahiri, S., Chen, S., Pan, R., Dillig, I., Born, C., Naseer, I.:
  Formal specification and verification of smart contracts for azure blockchain
  (April 2019),
  https://www.microsoft.com/en-us/research/publication/formal-specification-and-verification-of-smart-contracts-for-azure-blockchain

\bibitem{solidity}
Wood, G.: Solidity 0.6.8 documentation (May 2020),
  https://solidity.readthedocs.io/en/v0.6.8/

\bibitem{wood2014ethereum}
Wood, G., et~al.: Ethereum: A secure decentralised generalised transaction
  ledger. Ethereum project yellow paper  \textbf{151}(2014),  1--32 (2014)

\bibitem{wu2020ethscope}
Wu, L., Wu, S., Zhou, Y., Li, R., Wang, Z., Luo, X., Wang, C., Ren, K.:
  Ethscope: A transaction-centric security analytics framework to detect
  malicious smart contracts on ethereum. arXiv preprint arXiv:2005.08278
  (2020)

\bibitem{zhang2020txspector}
Zhang, M., Zhang, X., Zhang, Y., Lin, Z.: {TXSPECTOR}: Uncovering attacks in
  ethereum from transactions. In: 29th {USENIX} Security Symposium ({USENIX}
  Security 20). pp. 2775--2792. {USENIX} Association (Aug 2020)

\bibitem{firstparity}
Zhao, W.: {\$30 Million: Ether Reported Stolen Due to Parity Wallet Breach}
  (jul 2017),
  https://www.coindesk.com/30-million-ether-reported-stolen-parity-wallet-breach

\bibitem{zhou2020ever}
Zhou, S., Yang, Z., Xiang, J., Cao, Y., Yang, Z., Zhang, Y.: An ever-evolving
  game: Evaluation of real-world attacks and defenses in ethereum ecosystem.
  In: 29th {USENIX} Security Symposium ({USENIX} Security 20). pp. 2793--2810.
  {USENIX} Association (Aug 2020)

\end{thebibliography}

\end{document}